# A 4-Channel 10-Gbps/ch CMOS VCSEL Array Driver with on-chip Charge-pumps

**X. Huang,**[a,b] **D. Gong,**[b,1] **Q. Sun,**[b] **C. Chen,**[a,b] **D. Guo,**[a] **S. Hou,**[c] **G. Huang,**[a,1]
**S. Kulis,**[d] **C. Liu,**[b] **T. Liu,**[b] **P. Moreira,**[d] **A. Sánchez Rodríguez,**[d] **H. Sun,**[a,b] **J. Troska,**[d]
**L. Xiao,**[a] **L. Zhang,**[a,b] **W. Zhang,**[a,b] **and J. Ye**[b]

[a] *Central China Normal University,*
  *Wuhan, Hubei 430079, P.R. China*

[b] *Southern Methodist University,*
  *Dallas, TX 75275, USA*

[c] *Academia Sinica,*
  *Nangang, Taipei 11529, Taiwan,*

[d] *CERN,*
  *1211 Geneva 23, Switzerland*
  *E-mail*: dgong@smu.edu and gmhuang@mail.ccnu.edu.cn,

We present the design and test results of a 4-channel 10-Gbps/ch Vertical-Cavity Surface-Emitting Laser array driver, the cpVLAD. With on-chip charge-pumps to extend the biasing headroom for the VCSELs needed for low temperature operation and mitigation of the radiation effects. The cpVLAD was fabricated in a 65-nm CMOS technology. The test results show that the cpVLAD is capable of driving VCSELs with forward bias voltages as high as 2.8 V from a 2.5 V power supply. The power consumption of the cpVLAD is 94 mW/ch.

---

[1] Corresponding authors.

## 1. Introduction

The optical links for the high luminosity LHC (HL-LHC) upgrade are moving towards 10-Gigabit-per-second (Gbps) data rates per channel. Recent research indicates that the forward voltage of the Vertical-Cavity Surface-Emitting Laser (VCSEL) diodes significantly increases with radiation, especially at low temperatures [1]. These temperatures correspond to typical environment conditions for the inner track systems in HL-LHC. Considering that the power supply of the front-end system is restricted to 2.5 V, the increase of the VCSEL's forward voltages would finally result in the malfunction of the conventional VCSEL drivers due to the small voltage headroom at their output driving stage. The cpVLAD is a 4-channel 10-Gbps/ch VCSEL array driver with on-chip charge-pumps to boost the power supply of the driver stage so that it can adapt to different VCSEL forward voltage during the life cycle of the devices.

The cpVLAD receives differential signals with a minimum of 100 mV (P-P) from the lpGBT ASIC at 10 Gbps. The cpVLAD is capable of driving a common-cathode VCSEL array with a modulation current of no less than 6 mA and a bias current of no less than 5 mA. The size of cpVLAD is 1.835 mm x 1.635 mm with a pad frame compatible with the chosen VCSEL arrays. The cpVLAD is specified to operate within a temperature range of -35 $^{\circ}$C to +60 $^{\circ}$C and tolerate a total ionizing dose of 1 MGy.

## 2. cpVLAD Architecture

### 2.1 Overall structure

The block diagram of the cpVLAD is shown in Figure 1. Each channel includes a limiting amplifier (LA), an output driver, and a change-pump with a feedback circuit. All charge-pumps in the cpVLAD share a Voltage-Controlled Oscillator (VCO) and an I$^2$C slave module.

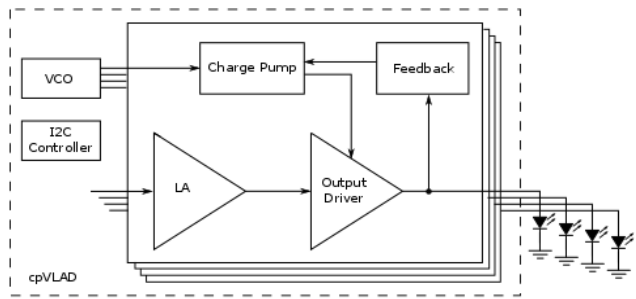

Figure 1. Block diagram of the cpVLAD.

### 2.2 Limiting Amplifier

The schematic of the LA is shown in Figure 2 (a). The LA consists of four cascaded gain stages with DC-coupled fully differential structures. In each stage, an inductive shunt peaking technique is adopted to extend the bandwidth. Center-tapped three-terminal inductors and shared inductor topology are used to reduce the layout area. The gain of the first two stages is 10.2 dB with a bandwidth of 11.3 GHz. The gain of the last two stages is 11.4 dB with a bandwidth of 14.2 GHz. Overall, the LA achieves a total gain of 21.5 dB and a bandwidth of 12.0 GHz with a power consumption of 15 mW.

### 2.3 Output Driver

The schematic of the output driver is shown in figure 2 (b). A differential pair $NM_1/NM_2$ works as a current switch. The currents $I_{mod}$ and $I_{tot}$ are programmable via I2C-controlled DACs. $R_1$ and $PM_1$ are the load of the left branch and behave as an active inductor to boost the bandwidth of the output driver. The bandwidth of the output driver for the typical process corner, 2.5 V, and 27 °C is 20.25 GHz.

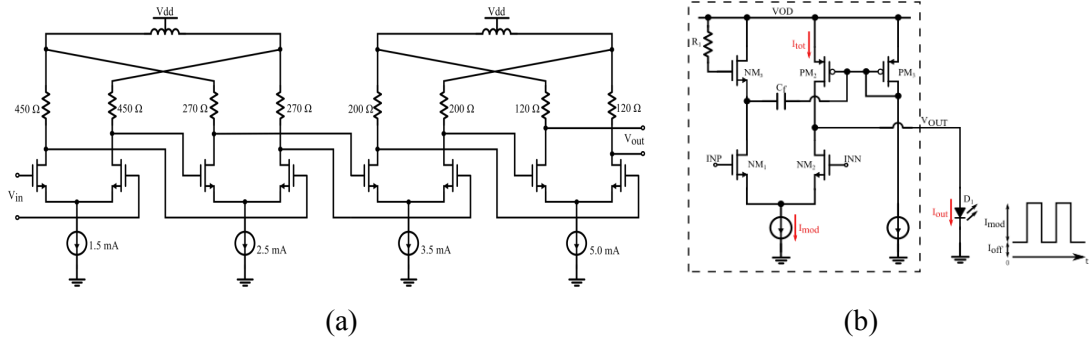

(a)                                                     (b)

Figure 2. Schematic of LA (a) and output driver (b).

### 2.4 Charge-Pump

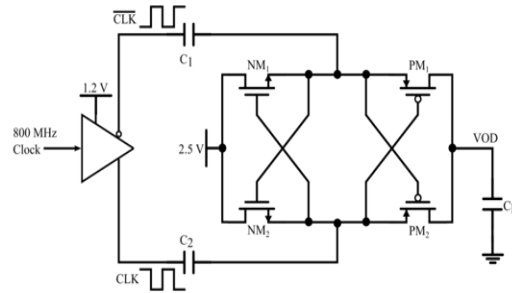

Figure 3. Schematic of charge-pump.

The schematic of the charge-pump circuit [2] is shown in figure 3. $C_1$ and $C_2$ are the booster capacitors, and $C_L$ is the decoupling capacitor. Four MOS transistors ($NM_1$, $NM_2$, $PM_1$, and $PM_2$) are used as switches. The cross-connected NMOS transistors $NM_1$ and $NM_2$ constitute the path to charge the booster capacitors $C_1$ and $C_2$, while the cross-connected PMOS transistors $PM_1$ and $PM_2$ establish the path to transfer the charge from $C_1/C_2$ to $C_L$. A VCO generates two complementary 800 MHz clock signals CLK and $\overline{CLK}$ to control the charge and discharge processes in turn. When CLK is high and $\overline{CLK}$ is low, $NM_1$ and $PM_2$ are on, while $NM_2$ and $PM_1$ are off. The capacitor $C_1$ is charged by the 2.5 V power supply, and disconnected from the output. Meanwhile, the charge on the capacitor $C_2$, is transferred to $C_L$ as the output. When CLK is low and $\overline{CLK}$ is high, $NM_2$ and $PM_1$ are on, while $NM_1$ and $PM_2$ are off. The capacitor $C_1$ discharges through $PM_1$ to boost the output. It should be noted that the clock signals CLK and $\overline{CLK}$ provide the DC power supply $V_{CLK}$ of 1.2 V to boost the output voltage. A negative feedback circuit is used to adjust automatically the charge-pump voltage VOD and keep the driver operating at an appropriate condition. In simulations, 30 ns after power up, the output voltage of the charge-pump settles to a steady voltage of approximately 3.2 V at the load current of 16 mA with a ripple of 60 mV (P-P). When the load current is from 6 mA to 18 mA, the efficiency range of the charge-pump is from 60% to 70%.

### 2.5  I$^2$C module and ASIC layout

The I$^2$C module is designed to configure the 26-byte internal registers of the cpVLAD. A triple modular redundancy (TMR) technique is used to protect the internal registers and the I$^2$C module from the radiation-induced single event upsets (SEUs). The cpVLAD was fabricated in a 65-nm CMOS technology. The die size is 1.835 mm x 1.635 mm.

### 3.  Test result

To characterize the cpVLAD, two boards were designed, one to characterize the ASIC electrically and the other when driving a VCSEL. The dies were wire-bonded to the test boards. Both the electrical and optical eye diagrams were measured at 10 Gbps with an input amplitude of 200 mV (P-P). Figure 6 (a) shows the optical eye diagram of Channel 1 with a 6.3 mA modulation current and a 2.1 mA offset current. The optical eye has passed the 10 Gigabit Ethernet (GbE) eye mask [3] test. The power consumption is 94 mW/ch. The VCSEL bias voltage headroom was measured by sourcing the output current into a programmable voltage supply. Sweeping its DC voltage allowed to emulate the VCSEL forward voltage between 1.6 and 3.2 V. The output modulation amplitude versus the emulated VCSEL forward voltage is shown in figure 4 (b). The cpVLAD can operate correctly with VCSEL forward voltages up to 2.8 V while still being able to properly modulate the VCSEL.

The cpVLAD was measured and performed according to the specification for the specified ambient temperature range. Devices mounted on the electrical characterization boards were irradiated by x rays for 69 hours to a total ionizing dose (TID) of 9 MGy with a dose rate of 130 kGy/h. Devices in the Optical characterization boards were irradiated for 90 hours to 6.3 MGy dose with a dose rate of 70 kGy/h. The electrical output modulation amplitude, with a simulated 2.4 V VCSEL bias voltage, and the optical modulation amplitude versus the TID are plotted in figure 7 (a) and 7 (b), respectively. The tested devices functioned normally after irradiation.

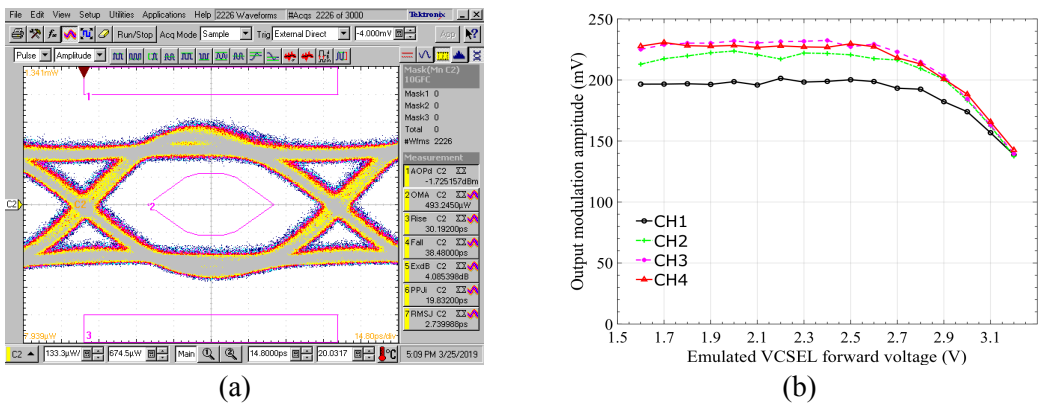

(a)                                   (b)

Figure 6. Optical eye diagram (a) and VCSEL forward voltage sweeping (b).

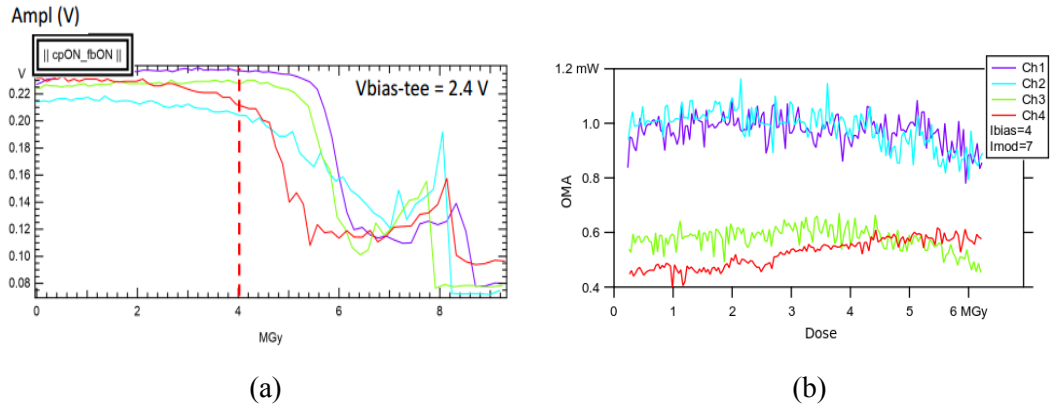

|   (a)   |   (b)   |
|---|---|

Figure 7. Electrical output modulation amplitude (a) and optical modulation amplitude (b) versus radiation dose under electrical irradiation.

## 4. Conclusion

A 4×10 Gbps VCSEL array driver, cpVLAD, was designed and fabricated in a 65-nm CMOS technology. To overcome the headroom limitation resulting from the low supply voltage and the high VCSEL forward voltage at low temperatures and after irradiation in the high energy physics experiments, on-chip charge-pumps were implemented to increase the power supply voltage of the output driver. The cpVLAD occupies an area of 1.835 mm x 1.635 mm. Both the electrical and optical tests have demonstrated that the cpVLAD is able to operate with a range of VCSEL forward voltage of up to 2.8 V and to withstand TID radiation doses up to 6.3 MGy. The power consumption of the cpVLAD is 94 mW/channel.

## Acknowledgments

This work is supported by the NSF and the DOE Office of Science, SMU's Dedman Dean's Research Council Grant, and the National Science Council in Taiwan. We want to thank James Kierstead from Brookhaven National Laboratory for the help in the irradiation test.